\documentclass[fleqn,10pt]{wlscirep}

\usepackage{amsmath}
\usepackage{amsfonts}
\usepackage{amssymb}

\title{Effects of temporal correlations in social multiplex networks} 

\author[1,2]{Michele Starnini}
\author[3]{Andrea Baronchelli}
\author[4,*]{Romualdo Pastor-Satorras}

\affil[1]{Departament de F\'{\i}sica Fonamental, Universitat de
  Barcelona, Mart\'{\i} i Franqu\`es 1, 08028 Barcelona, Spain}

\affil[2]{Universitat de Barcelona Institute of Complex Systems
  (UBICS), Universitat de Barcelona, Barcelona, Spain}

\affil[3]{Department of Mathematics - City, University of London -
  Northampton Square, London EC1V 0HB, UK}

\affil[4]{Departament de F\'\i sica, Universitat Polit\`ecnica de
  Catalunya, Campus Nord B4, 08034 Barcelona, Spain}

\affil[*]{romualdo.pastor@upc.edu}

\begin{abstract}
  Multi-layered networks represent a major advance in the description of
  natural complex systems, and their study has shed light on new
  physical phenomena.  Despite its importance, however, the role of the
  temporal dimension in their structure and function has not been
  investigated in much detail so far.  Here we study the temporal
  correlations between layers exhibited by real social multiplex
  networks. At a basic level, the presence of such correlations implies a certain degree
  of predictability in the contact pattern, as we quantify by an extension of the entropy and
  mutual information analyses proposed for the single-layer case.  At a
  different level, we demonstrate that temporal correlations are a signature
  of a `multitasking' behavior of network agents, characterized by a
  higher level of switching between different social activities than
  expected in a uncorrelated pattern. Moreover, temporal correlations
  significantly affect the dynamics of coupled epidemic processes
  unfolding on the network.  Our work opens the way for the systematic
  study of temporal multiplex networks and we anticipate it will be of
  interest to researchers in a broad array of fields.
\end{abstract}

\begin{document}

\flushbottom
\maketitle

\thispagestyle{empty}

\section*{Introduction}

Recently, the theoretical framework of network science~\cite{Newman2010}
has been enriched by two new concepts: \textit{Multiplex
  networks}~\cite{Boccaletti20141,Kivela2013,Lee2015}, whose edges
belong to different layers, representing different kinds of
interactions; and \textit{temporal
  networks}~\cite{Holme:2011fk,Holme2015}, whose edges have an intrinsic
dynamics of creation and annihilation, representing interactions
switching on and off with given characteristic time scales. The
introduction of these two viewpoints has greatly enrich our
understanding of real networks.  On the one hand, the multiplex
representation, through the definition of new observables, such as
multilayer clustering, degree correlations or layer overlap
\cite{Boccaletti20141}, has allowed for a better structural
characterization of many networked systems, and helped clarify the
behavior of dynamical processes on top of
them~\cite{De-Domenico:2016aa,citeulike:13204123,Buldyrev2010,6517108,Dickison2012}.
On the other hand, taking into account the temporal dimension of edges
has allowed to uncover unexpected properties of time-varying networks,
such as their general bursty nature, characterized by a heavy-tailed
distribution of inter-event times $\tau$ between the establishment of
consecutive connections~\cite{barabasi2005origin,Stehle:2011}, often
compatible with power-law forms, $\psi(\tau) \sim
\tau^{-(1+\alpha)}$. These temporal effects, moreover, have been shown
to radically alter the behavior of dynamical processes on such evolving
structures~\cite{dynnetkaski2011,PhysRevLett.98.158702,Parshani:2010,PhysRevE.94.022316}.

In the particular case of social networks~\cite{Jackson2010}, the recent
availability of large digital databases has shown the necessity of a
dual description based on both multiplex and temporal network
approaches.  This urgency stems from the very nature of social
interactions, which are diverse in nature and quality, with different
layers co-existing and interacting with one another (e.g., physical
vs.~digital interactions)~\cite{Verbrugge01061979}, and evolve in time,
with new relationships being continuously created and destroyed.
Therefore, a realistic description should rely on \textit{temporal
  multiplex networks},
that can be mathematically described by
endowing the multiplex paradigm with an additional temporal dimension,
see Methods.
The empirical evidence of this dual nature of social networks is
arousing a growing interest in their temporal multiplex representation
within the complex system
community~\cite{Vijayaraghavan,PhysRevLett.111.058702,PhysRevLett.111.058701}.
However, the effects of the interplay between temporal and multiplex
dimensions on the structure and function of real networks still remain
largely unexplored, also due to the lack of suitable, longitudinal data.

In this paper, we will consider one particular aspect, namely the
possibility of observing correlations between the temporal activity of
different layers. In single-layered networks, indeed, temporal
correlations have been recently observed~\cite{Karsai:2012aa}, implying
the presence of memory effects~\cite{PhysRevE.92.022814}.  In the
context of temporal multiplex networks, this effect translates into the
possible presence of \textit{inter-layer temporal correlations},
i.e.~the fact that a social interaction, taking place in some given
layer at some given time, might alter the probability of subsequent
interactions in different layers.  Such correlations have been
characterized in Ref.~\cite{Vijayaraghavan} in terms of a Multiplex
Markov chain, showing the presence of correlated creation and
destruction of connections between pairs of nodes in different
layers. Here we focus on the effects of such temporal correlations, both
in the dynamics of social interactions and on dynamical processes
running on top of a temporal multiplex. We start by checking the
presence of inter-layer temporal correlations in several empirical
scenarios, by applying a simple information theory approach, which
reveals a certain degree of potential predictability in the interaction
patterns.  We measure the effect of temporal correlations on social
activity by defining a multitasking index, and show that they tend to
increase the rate of switching between layers expected in an
uncorrelated setting.  Finally, we explore the impact of temporal
correlations on the dynamics of coupled epidemic/awareness processes
unfolding on different layers~\cite{Granell2013}, showing that they can
either slow down or speed up the epidemic spread, depending on the
region of the parameter space defining the model.

In our analysis, we consider different empirical scenarios: Human
contact networks, recorded by the ``Reality Mining'' (RM)
experiment~\cite{eagle2006reality} and consisting of two independent
data sets, ``Social Evolution'' (SE) and ``Friends and Family'' (FF);
Open Source Software (OSS) collaboration
networks~\cite{PhysRevE.91.052813}, with data provided by a OSS project
part of the Apache software foundation~\cite{Apache}; and scientific
collaboration networks~\cite{newmancitations01}, reconstructed from the
American Physical Society (APS) data sets for research~\cite{apsdata}.
In all cases, interactions are represented as a temporal multiplex
network formed by two layers, arbitrarily denoted $\ell = 1$ and
$\ell = -1$.  See Methods and Section 1 of the Supplementary Material
for a full description of the considered data sets.

\section*{Results}

\subsection*{Correlation and influence between layers}

One simple approach to establish the presence of inter-layer temporal
correlations in our empirical datasets consists in extending to
multiplex networks the mutual information analysis traditionally used to
detect temporal correlations in single layer sequences of social
activity~\cite{Song1018,PhysRevX.1.011008,mobilitypetri2012}.  In
multiplex networks, an individual $i$ switching from one kind of
interaction to another one (e.g.~he/she sends an email to a colleague
and then co-edits some code with another collaborator) is represented by
a link between node $i$ and node $j$ in layer $\ell$ at time $t_1$ and a
link between node $i$ and node $k$ (including the case $j=k$) in layer
$-\ell$ at time $t_2 > t_1$.  We want to understand whether $i$, after
having an interaction with $j$ in layer $\ell$, chooses his next partner
$k$ in layer $-\ell$ at random or there is a certain degree of
predictability in his choice~\cite{Song1018,PhysRevX.1.011008}.

To address this issue, we define the uncorrelated entropy $H^u_i(\ell)$
of individual $i$ as
\begin{equation}
  \label{eq:h1}
  H^u_i(\ell) = - \sum_{j_\ell} p_i (j_\ell) \ln[ p_i ( j_\ell)] , 
\end{equation}
where $p_i ( j_\ell)$ is the probability that individual $i$ interacts
with individual $j$ in layer $\ell$.  The uncorrelated entropy thus
measures the degree of heterogeneity in the interactions pattern of an
individual in one layer.  The conditional entropy
$H^c_i(\ell \rightarrow -\ell)$ of individual $i$ from layer $\ell$ to
layer $-\ell$ is defined as
\begin{equation}
  \label{eq:h2}
  H^c_i(\ell \rightarrow -\ell) = - \sum_{j_{\ell}} p_i(j_{\ell})
  \sum_{k_{-\ell}}  
  p_i (k_{-\ell} | j_{\ell}) \ln[  p_i (k_{-\ell} | j_{\ell}) ], 
\end{equation}
where $p_i (k_{-\ell} | j_{\ell})$ is the conditional probability that
individual $i$ interacts with individual $k$ in layer $-\ell$
immediately after interacting with individual $j$ in layer $\ell$.  
The influence of layer $\ell$ on layer $-\ell$ 
is quantified by the mutual information, defined for each individual $i$ as
the difference between uncorrelated and conditional entropy,
$I_i(\ell \rightarrow -\ell) = H^u_i(-\ell) - H^c_i(\ell \rightarrow
-\ell)$, thus
\begin{equation}
\label{eq:diff_h}
I_i(\ell \rightarrow -\ell) =  \sum_{j_{\ell}, k_{-\ell}} p_i
(j_{\ell}, k_{-\ell})  
\ln \left( \frac{p_i (j_{\ell}, k_{-\ell})}{p_i (j_{\ell}) p_i(
    k_{-\ell})}   \right),  
\end{equation}
where $p_i ( k_{-\ell}, j_{\ell} )$ is the joint probability that
individual $i$ interacts first with individual $k$ in a layer $-\ell$
and immediately after with individual $j$ in a layer $\ell$.  Since
$H^u_i \geq H^c_i$, the mutual information $I_i$ is always positive, and
it is equal to zero only if the interaction patterns of individual $i$
on the two layers $-\ell$ and $\ell$ are temporally uncorrelated.
Therefore, $I_i(\ell \rightarrow -\ell)$ measures the degree of
potential predictability of the interaction pattern of individual $i$ in
layer $-\ell$, and it is equal to the amount of information about his
next partner in layer $-\ell$ earned by knowing his current partner in
layer $\ell$. 

To avoid spurious effects due sample size issues, in the computation of
these quantities we perform a bootstrap analysis, retaining only those
individuals who have a value of the conditional entropy significantly
smaller than the one obtained by rewiring the network according to a
null model, in which, for each individual $i$, the set of all pairs
consecutive interactions in different layers is extracted, and the set
of second individual interactions in each pair is randomized. This
procedure destroys any temporal correlations between layers, while
keeping constant the uncorrelated entropy. 
See Section 2 of the Supplementary Material for further details.

\begin{figure}[t]
  \centering
  \includegraphics*[width=0.75\columnwidth]{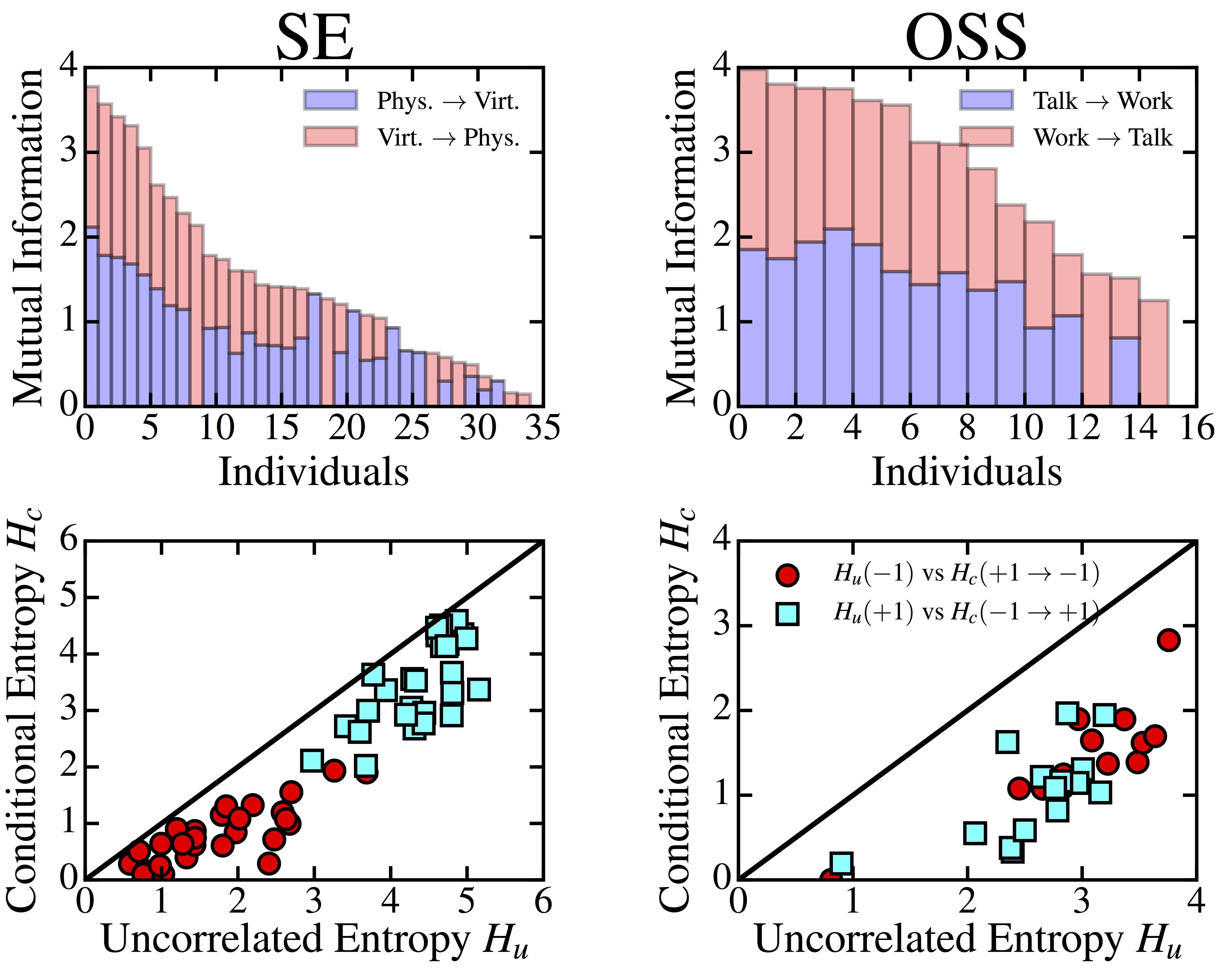}
  \caption{Scatter plot of uncorrelated vs conditional entropy of each
    individual $i$, $H^u_i(\ell)$ vs $H^c_i(\ell \rightarrow -\ell)$
    (bottom row), and mutual information between layers,
    $I_i(\ell \rightarrow -\ell)$ (top row).  Only individuals with a
    conditional entropy with a $p$-value smaller than $0.05$ with
    respect to the null model are plotted.  Data shown are: RM contact
    network, data set SE (left) and OSS collaboration network (right). }
   \label{fig:entropy}
\end{figure}

Fig.~\ref{fig:entropy} (bottom panels) shows the relation between
uncorrelated and conditional entropy for single individuals, on the SE
contact and OSS collaboration networks (see Supplementary Fig.~S2 for
additional datasets).  One can see that many individuals show a
significant entropy difference, resulting in a certain degree of
potential predictability, in each data set under consideration.  For the
case of RM contact networks, in both data sets SE and FF, the
uncorrelated and conditional entropy obtained in the physical layer
($\ell = +1$) are larger than the ones obtained in the digital layer
($\ell = -1$), because the former is characterized by a richer pattern
of interactions, with a larger density and heterogeneity (see
Supplementary Table S1).  The same behavior is observed in the OSS
collaboration network, where the denser communication layer
($\ell = -1$) shows larger values of the uncorrelated and conditional
entropy than the ones obtained in the co-work layer ($\ell = +1$).
Fig.~\ref{fig:entropy} (top panels), see also Supplementary Fig.~S2,
shows the amount of potential predictability of an individual $i$ in one
layer $\ell$ obtained by the other layer $-\ell$, as measured by mutual
information $I_i(\ell \rightarrow -\ell)$.  For the majority of
individuals there is a mutual influence between layers, both
$I(+1 \rightarrow -1) > 0$ and $I(-1 \rightarrow +1) > 0$.

These results show, in agreement with Ref.~\cite{Vijayaraghavan}, that
the sequences of contacts in social multiplex networks present indeed
temporal correlations. In our mutual information approach, these
correlations translate in a certain degree of
predictability~\cite{Song1018,PhysRevX.1.011008}, resulting from a
deterministic component that overrules the random establishment of
contacts in one or another social layer. In the following we will show
how these temporal correlations can have an impact on social behavior
and dynamic spreading.

\subsection*{Multitasking index of individuals}

The temporal correlations observed in the entropy analysis performed
above have an effect in the patterns of social interactions that can be
gauged by using simple observables.  Considering the number of
interactions, $n_{\ell}^{(i)}(\Delta t)$ and
$n_{-\ell}^{(i)}(\Delta t)$, that an individual $i$ performs in a time
interval $\Delta t$ in two different layers $\ell$ and $-\ell$, a
\textit{multitasking index} $r_{i}(\Delta t)$ of individual $i$ can be
defined as the Pearson correlation coefficient between the set of
variables $\{ n_{\ell}^{(i)}(\Delta t), n_{-\ell}^{(i)}(\Delta t) \}$,
where each pair $(n_{\ell}(\Delta t), n_{-\ell}(\Delta t) )$ is measured
at different time intervals of fixed length $\Delta t$.  If
$r_{i}(\Delta t)>0$ (i.e. if the values of $n_{\ell}^{(i)}(\Delta t)$
and $n_{-\ell}^{(i)}(\Delta t)$ attain comparable values), then
individual $i$ is simply distributing his activity among the two layers
and he/she is likely to interact indistinctly in both layers at the same
time.  Otherwise, if $r_{i}(\Delta t) < 0$ (i.e. if a large
$n_{\ell}^{(i)}(\Delta t)$ is associated with a small
$n_{-\ell}^{(i)}(\Delta t)$, and vice-versa), then he/she is likely to
be concentrating her activity on one of the two layers.

\begin{figure}[t]
  \centering
  \includegraphics*[width=0.75\columnwidth]{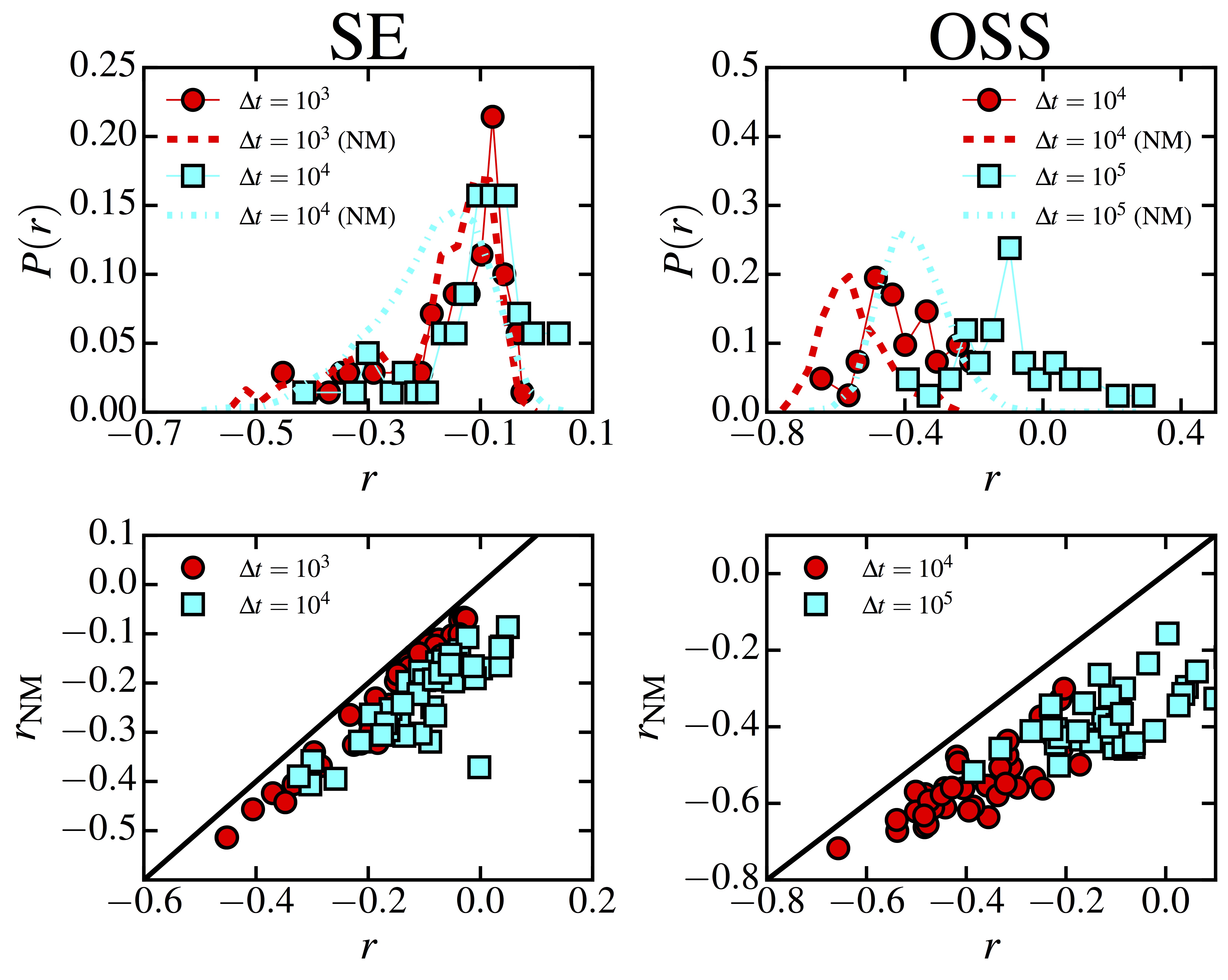}
  \caption{Comparison of the multitasking index of individuals in the
    original data, $r$, with the corresponding index $r_{NM}$ in data
    randomized according the null model.  Probability distribution of
    the multitasking index of the original and randomized data, $P(r)$
    and $P(r_{NM})$ (top row), and scatter plot of the multitasking
    index of the original versus randomized data, $r$
    vs. $r_\mathrm{NM}$ (bottom row), for different time window
    $\Delta t$.  In the scatter plots, only individuals with $r$ with a
    p-value smaller than 0.05 or greater than 0.95 with respect to the
    null model are plotted.  In calculating the multitasking index, we
    consider only individuals with at least $10$ interactions in each
    layer. Data shown are: RM contact network, data set SE, 41
    significant individuals over 73 (left, $\Delta t$ expressed in
    seconds) and OSS collaboration network, 40 significant individuals
    over 52 (right, $\Delta t$ expressed in minutes).}
  \label{fig:pearson}
\end{figure}

Fig.~\ref{fig:pearson} (top row) shows the probability distribution of
the multitasking index, $P(r)$, measured for each node of the SE contact
and OSS collaboration networks (see Supplementary Fig.~S3 for additional
datasets), for different values of the time interval $\Delta t$,
obtained by cutting the whole temporal sequence into consecutive slices.
The multitasking index is generally negative, indicating the presence of
large sequences of uninterrupted acts of the same kind.  In a given time
interval $\Delta t$, an individual is more likely to relate with the
others only through face-to-face interactions, or only through calls or
texts, and less likely to use both channels simultaneously.  In the
context of the OSS collaboration network, this translates into
developers being more likely either to communicate or to co-work, not
doing both actions at the same time.  In APS networks, see Supplementary
Fig.~S3, it implies that authors are more likely to collaborate in a
sequence of papers in the same journal, instead of switching among
different journals.

This apparently strong effect must be, however, judged with caution,
since burstiness alone is a sufficient condition for the emergence of
large sequences of consecutive interactions in the same layer, even in
temporally uncorrelated networks.  In Section 4 of the Supplementary
Material, indeed, we build a null model of an uncorrelated temporal
multiplex network with two independent renewal processes, one for each
layer $\ell$, each one with a power-law form of the inter-event time
distribution, $\psi_{\ell}(\tau) \sim \tau^{1- \alpha_{ell}}$.  We
analytically show that the probability distribution of finding $n$
consecutive events on the same layer $\ell$ follows a power-law form,
$P_{\ell}(n) \sim n^{-(1 + \alpha_{-\ell}/\alpha_{\ell}) }$, see
Supplementary Figs. S5 and S6.

Therefore, the empirical multitasking index needs to be contrasted with
a null model which destroys temporal correlations between layers. In
this null model, for each individual $i$, the set all its interactions
in each layer is randomized, in such a way that the interevent time
distribution $\psi_\ell(\tau)$ of each layer $\ell$ is preserved, while
temporal inter-layer correlations are destroyed.  In
Fig.~\ref{fig:pearson}, upper row, we show also the distribution of
multitasking indexes in the randomized versions of our empirical
datasets. As we can see, the real and randomized distributions are
clearly different, specially for larger values of the time interval
$\Delta t$.  In Fig.~\ref{fig:pearson}, bottom row (see also
Supplementary Fig.~S3), we present a scatter plot between the empirical
and randomized coefficients, $r(\Delta t)$ and $r_{NM}(\Delta t)$,
respectively, for different time intervals $\Delta t$.  Only individuals
whose coefficient $r(\Delta t)$ is significantly different from
$r_{NM}(\Delta t)$ (with a $p$-value smaller than $0.05$ or greater than
$0.95$, see Section 2 of the Supplementary Material, are plotted.

One can see that almost all significant individuals have a multitasking
coefficient $r(\Delta t)$ greater than the corresponding coefficient
$r_{NM}(\Delta t)$ obtained in the null model, as highlighted by the
diagonal line.  This implies that, in general, temporal inter-layer
correlations tend to decrease the stretches of time in which activity is
concentrated in a single layer, increasing the multitasking index.
The practical implication of this observation for social dynamics is that 
temporal correlations increase the rate at which
individuals switch from one kind of social activity from another one,
with respect to a purely random behavior, 
only constrained by the burstiness of human dynamics.

\subsection*{Effects of temporal correlations on coupled spreading dynamics}

As we have seen, temporal correlations can alter the pattern of social
interactions. Additionally, they can also influence the behavior of
dynamical processes running on top of temporal multiplex networks. To
show this, we consider the interplay of competing spreading processes,
which has been previously studied on static, synthetic, multiplex
networks \cite{De-Domenico:2016aa, Granell2013}.  In this framework, an
epidemic spreads on the physical layer of the RM contact networks while
information spreads on the virtual layer, representing awareness to
prevent the infection \cite{Granell2013}. This scenario is modeled as
follows: A Susceptible-Infected process runs on the physical layer, in
which whenever an infected ($I$) individual $i$ has a contact with a
susceptible ($S$) one $j$, the disease is transmitted with probability
$\beta_1$, and $j$ becomes infected.  An Unaware-Aware process runs on
the virtual layer, in which whenever an aware ($A$) individual $i$ has a
contact with an unaware ($U$) one $j$, the information is transmitted
with probability $\beta_2$, and $j$ becomes aware.  Infected individuals
are instantaneously aware of the disease, while a susceptible individual
that becomes aware of the disease is instantaneously immunized ($R$)
from it, and cannot be infected.

\begin{figure}[t]
  \centering
\includegraphics*[width=0.75\columnwidth]{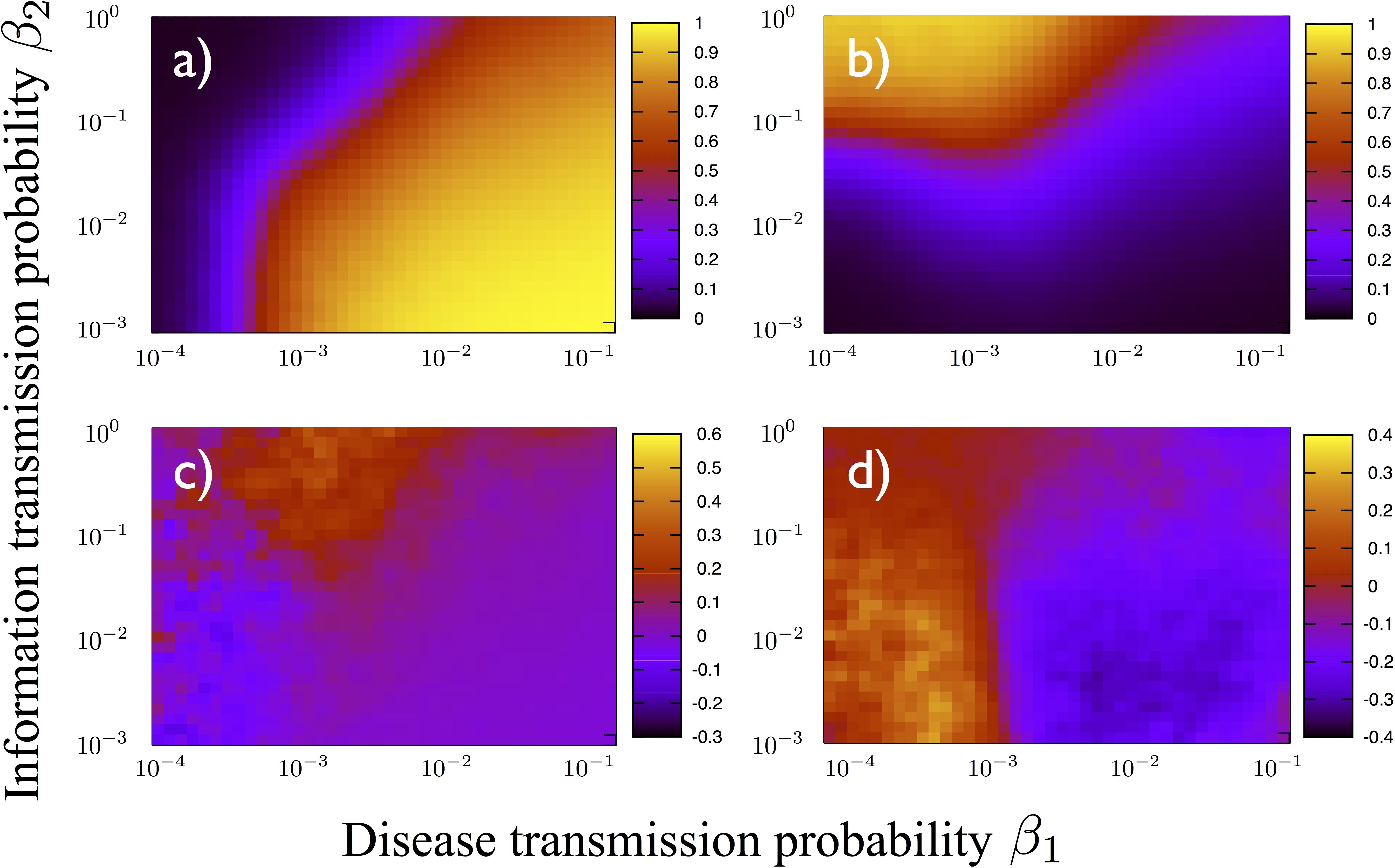}
\caption{Phase diagrams ($\beta_1,\beta_2$) obtained by simulating the
  competition between epidemic spreading and information awareness on
  the RM contact network, data set SE (see Supplementary Fig.~S4 for
  data set FF).  Plots show (\textbf{a}) the fraction of infected and
  (\textbf{b}) fraction of immunized individuals for the original data;
  relative difference of infected (\textbf{c}) and immunized
  (\textbf{d}) individuals with respect to randomized data. }
   \label{fig:epidemics}
\end{figure}

Fig. \ref{fig:epidemics} (top row) shows the final prevalence
$\rho = I_{\inf}/N$ (\textbf{a}) and the fraction of immunized
individuals $i = R_{\inf}/N$ (\textbf{b}) measured at the end of the
contact sequence of data set SE (see also Section 3 of Supplementary
Material and Supplementary Fig.~S4 for data set FF), as a function of
the two parameters $\beta_1$ and $\beta_2$ controlling the dynamics.
The population shows a clear transition from an inactive
(i.e. susceptible) to an active (i.e. infected) state, for increasing
values of the infection probability $\beta_1$, and decreasing values of
the probability of informqation transmission $\beta_2$.  Interestingly,
the fraction of immunized agents does not follow such behavior with
respect to $\beta_1$, but it reaches a maximum for
$\beta_1 \simeq 10^{-3}$, and decreases for larger values.  The effects
of temporal correlations are shown in the bottom row of
Fig. \ref{fig:epidemics}, where we plot the relative prevalence
$\rho_R = (\rho_\mathrm{NM} - \rho)/\rho$, (\textbf{c}) and relative
fraction of immunized individuals, $i_R = (i_\mathrm{NM} - i)/i$,
(\textbf{d}), as obtained by contrasting original data with a null model
(NM) in which the times of the sequence of contacts between any given
pair of individuals $i$ and $j$ is randomized, destroying inter-layer
temporal correlations while keeping the inter-event time distribution of
contacts between pairs; see Section 2 of the Supplementary Material for
further details.

The effect of temporal correlations on the epidemic outbreak is complex
and nonlinear.  On the one hand, the coupled spreading processes
unfolding on a uncorrelated network result in a final prevalence up to
$50\%$ larger than the corresponding processes on a correlated
multiplex.  The maximum effect of temporal correlations on the
prevalence is observed for large $\beta_2$, close to the transition
between the inactive and active phases.  Therefore, temporal
correlations slow down the epidemic spreading in these regions of the
phase space, consequently reducing the disease outbreak.  On the other
hand, the final number of immunized individuals is larger in the
uncorrelated case with respect to the correlated one for small
$\beta_1$, while it is smaller for large $\beta_1$.  This implies that
temporal correlations slow down the information diffusion for small
$\beta_1$, and they speed it up for large $\beta_1$.

\section*{Discussion}

Here we have shown that the presence of temporal correlations between
the layers of a social multiplex networks can affect both the patterns
of social contacts and the behavior of unfolding spreading processes.
On the one hand, inter-layer correlations reduce the inclination of
individuals to engage in large sequences of interactions of the same
kind, as captured by the increase of their multitasking index. 
This observation means that individuals tend to switch from 
one kind of social activity to another one 
more frequently than would be expected in a purely random pattern of
interactions.  At the same level, this correlated pattern implies a
certain degree of predictability in the sequence of contacts.  On the
other hand, temporal correlations alter the dynamics of coupled
epidemic/awareness processes unfolding on different layers, either
enhancing or depressing the spreading speed.  In order to single out
genuine temporal correlations between layers, in our analysis we
contrast our results with appropriate null models, pointing out that the
burstiness of human activity within a single layer is responsible for
spurious correlations, and therefore it should be taken into account in
the definition and measurement of new quantities related with social
dynamics. Interestingly, the results obtained are independent of the
length of the temporal sequence defining the multiplex, as evidences by
the SE and FF datasets, resulting from similar experiments but with
widely different length.

Our study allows for a better understanding of social networks,
highlighting the interplay between their twofold temporal and
multi-layer nature, which allows to define and measure new observables
able to characterize the entanglement in the development of different
kinds of social activity.  Moreover, our findings pave the way to sense
and measure temporal correlations in other fields of complexity invested
by the multiplex representation, ranging from the multilayer
organization of brain networks~\cite{Bullmore2012} to multimodal
mobility and efficient transportation~\cite{Gallotti:2014aa,
  Gallottie1500445}, as well as to their extension to more general
multilayer networks.  In particular, further research is in order to
fully unravel the influence of inter-layer correlations in more complex
epidemic spreading processes, as well as their impact on immunization
strategies levering on the temporal patterns of social
interactions~\cite{Lee:2010fk,2013arXiv1305.2357S}.

\section*{Materials and Methods}

\subsection*{Mathematical description of temporal multiplex networks}

\begin{figure}[t]
  \begin{center}
    \includegraphics*[width=0.75\textwidth]{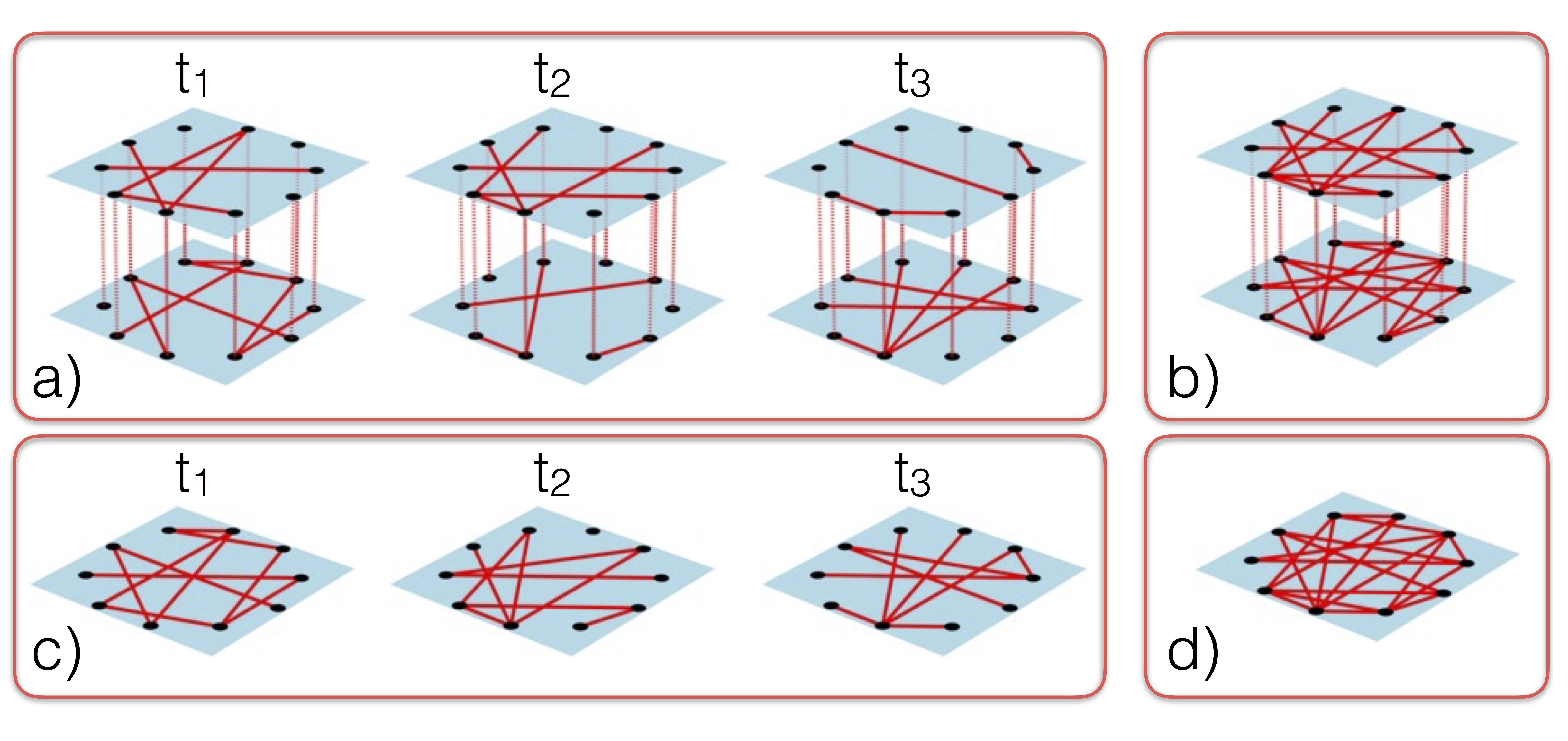}
  \end{center}
  \caption{Different observation levels of a temporal multiplex network.
    A full temporal multiplex network (\textbf{a}), in this case with
    two levels, is represented by different snapshots at times
    $t_i \in \mathcal{T} = \{t_1, t_2, t_3\}$ of a single set of nodes
    with edges on different layers (colors) that appear at different
    times. The integrated static multiplex (\textbf{b}) is given by the
    projection over the time window $\mathcal{T}$ of all edges, which
    appear in their respective layers if they have appeared at least
    once in the whole observation window. A single layer temporal
    network (\textbf{c}) is obtained by projecting all layers onto a
    single one. Simultaneous projection over time and layers leads to a
    single layer static network (\textbf{d}).}
  \label{fig:artwork}
\end{figure}

Temporal multiplex networks can be mathematically described by endowing
the multiplex paradigm with an additional temporal dimension
\cite{Vijayaraghavan}.  In this way, a temporal multiplex network can be
represented by a \textit{contact sequence}, a set of quadruplets
$(i,j,t,\ell)$ indicating that nodes $i$ and $j$ are connected at time
$t$ in layer $\ell$, with $i,j \in \mathcal{V} = \{1, \ldots, N \}$, the
set of nodes, of a total number $|\mathcal{V}| = N$, $t \in \mathcal{T}$
the set of contact times, and
$\ell \in \mathcal{L} = \{\ell_1, \ell_2, \ldots, \ell_L \}$, the set of
$|\mathcal{L}| = L$ layers.  From this exact description, coarse grained
information can be obtained by projecting either temporal, multiplex or
both dimensions onto a static and/or single-layered network, see
Fig. \ref{fig:artwork}.  A single-layered temporal network is obtained
by projecting different layers $\ell$ onto a single aggregate layer for
each contact time $t \in \mathcal{T}$, so that the resulting temporal
network is described in terms of a contact sequence with triplets
$(i,j,t)$.  A static multiplex network is recovered by projecting time
$t$ onto a time-aggregated network for each layer $\ell$, resulting in a
set of $L$ (possibly weighted) networks,
$\vec{G}=(G_{\ell_1},G_{\ell_2}, \ldots, G_{\ell_L})$.  Each network
$G_{\ell}$ is described by the adjacency matrix \cite{Newman2010}
$\textbf{a}^{\ell}$, whose elements
$a_{ij}^{\ell} = w_{ij}^{\ell} = \sum_{t} \chi(i,j,t,\ell)$ represent
the number of interactions between $i$ and $j$ occurring over the whole
contact sequence in layer $\ell$.  One can project both time $t$ and
multiplexity $\ell$ onto a time-aggregated, single-layered network $G$.
The elements of its adjacency matrix
$a_{ij}= w_{ij} = \sum_{\ell, t} \chi(i,j,t,\ell)$ represent the number
of interactions between $i$ and $j$ occurring over the whole contact
sequence across any layer $\ell$. The temporal dimension can also be
considered in more general multi-layer networks\cite{Boccaletti20141},
in which each layer is characterized by a different set of nodes.

\subsection*{Empirical data.}

We consider three different kinds of empirical temporal multiplex
networks, all formed by two layers (duplex): human contact networks,
recorded by the RM experiment \cite{eagle2006reality}, OSS collaboration
networks, reconstructed by means of data provided by the Apache software
foundation \cite{Apache}, and scientific collaboration networks,
reconstructed from the APS data set for research \cite{apsdata}.  The RM
experiment \cite{eagle2006reality}, conducted by the MIT Media Lab, is
composed by two data sets: ``Social Evolution'' (SE) and ``Friends and
Family'' (FF).  It records proximity data by means of bluetooth sensors,
forming a layer of physical interactions, $\ell=+1$, and digital
communications, as given by phone calls and text messages, merged in a
single layer of digital interactions $\ell=-1$.  The Apache software
foundation \cite{Apache} provides data of email communications between
developers and their commits to edit files of several OSS project.  We
focus on ``Apache Axis2/Java'', one of the project involving the largest
number of developers, and consider a layer of co-work, $\ell=+1$, formed
by co-commits to edit the same file, and a layer of communication,
$\ell=-1$, formed by email messages.  The APS dataset \cite{apsdata}
provides information about all papers published by the APS since 1893.
A multiplex network can be constructed by considering the co-authorship
of a paper published in any of the APS journals.  We consider a layer
formed by co-authorship in the journal Physical Review Letters (PRL),
$\ell=+1$, and coauthorship in other APS journal, excluding PRL,
$\ell=-1$.

\subsection*{Null models of temporal multiplex networks}

In order to single out inter-layer correlations in temporal multiplex
networks, we consider different null models. From a theoretical point of
view, the structure of a temporal multiplex network can be represented
as a collection of point processes \cite{cox1980point} for each layer
$\ell$, with two different levels of description:
\begin{enumerate}
\item A set of $N$ point processes,
  $\{ p_{\ell,i} \}_{i \in \mathcal{V}}$, where $\mathcal{V}$ is the set
  of layers, in which a point corresponds to an interaction of an agent
  $i$ with any other agent in the same layer;
\item A set of $N^2$ point processes,
  $\{ p_{\ell,i,j} \}_{i,j \in \mathcal{V}}$, in which a point
  corresponds to an interaction of agent $i$ with agent $j$ in the same
  layer.
\end{enumerate}
The simplest characterization of these point processes is in terms of
their inter-event time distributions representing the probability that
two points in a process are separated by a time $\tau$.  Therefore, a
null model of an uncorrelated temporal multiplex network corresponds to
$N \times L$ (or $N^2 \times L$) uncorrelated renewal processes
\cite{renewal}, depending on the level of coarse-graning one chooses to
consider, in which the time $\tau$ between two points is an independent
random variable distributed according to the inter-event time
distribution $\psi(\tau)$ extracted from the data.

From an empirical point of view, null models can be constructed from the
real data by a randomization processes \cite{Holme:2011fk}, in which
interactions in each layer are reshuffled, preserving certain physical
observables (mainly the inter-event time distributions).  The null model
for the multitasking index of individuals is based on the description
level (1), preserving the individual inter-event time distributions,
while case (2) has been used for the coupled spreading processes
unfolding on the multiplex network, preserving now the inter-event time
distributions of pairs of individuals. For the mutual information
analysis, a rewiring preserving the uncorrelated entropy was
performed. See Section 2 of the Supplementary Material for a detailed
definition of each empirical null model.

\section*{Acknowledgements}

We thank Professor Vladimir Filkov for sharing with us data of the Open
Source Software networks.  M.S. acknowledges financial support from the
James S. McDonnell Foundation.  R.P.-S. acknowledges financial support
from the Spanish MINECO under projects No.  FIS2013-47282-C2-2 and
FIS2016-76830-C2-1-P, EC FET-Proactive Project MULTIPLEX (Grant
No. 317532), and ICREA Academia, funded by the Generalitat de Catalunya.

\section*{Author contributions statement}

M.S., A.B., and R.P.-S. designed the research; M.S. performed the
analysis; M.S., A.B., and R.P.-S. analyzed the results and wrote the
paper.

\section*{Competing Interests}

The authors declare that they have no competing financial interests.


\begin{thebibliography}{10}
\expandafter\ifx\csname url\endcsname\relax
  \def\url#1{\texttt{#1}}\fi
\expandafter\ifx\csname urlprefix\endcsname\relax\def\urlprefix{URL }\fi
\expandafter\ifx\csname doiprefix\endcsname\relax\def\doiprefix{DOI }\fi
\providecommand{\bibinfo}[2]{#2}
\providecommand{\eprint}[2][]{\url{#2}}

\bibitem{Newman2010}
\bibinfo{author}{Newman, M. E.~J.}
\newblock \emph{\bibinfo{title}{Networks: An introduction}}
  (\bibinfo{publisher}{Oxford University Press}, \bibinfo{address}{Oxford},
  \bibinfo{year}{2010}).

\bibitem{Boccaletti20141}
\bibinfo{author}{Boccaletti, S.} \emph{et~al.}
\newblock \bibinfo{journal}{\bibinfo{title}{The structure and dynamics of
  multilayer networks}}.
\newblock {\emph{\JournalTitle{Phys. Rep.}}}
  \textbf{\bibinfo{volume}{544}}, \bibinfo{pages}{1 -- 122}
  (\bibinfo{year}{2014}).

\bibitem{Kivela2013}
\bibinfo{author}{Kivel\"{a}, M.} \emph{et~al.}
\newblock \bibinfo{journal}{\bibinfo{title}{{Multilayer Networks}}}.
\newblock {\emph{\JournalTitle{J. Complex Networks}}}
  \textbf{\bibinfo{volume}{2}}, \bibinfo{pages}{203--271}
  (\bibinfo{year}{2014}).

\bibitem{Lee2015}
\bibinfo{author}{Lee, K.-M.}, \bibinfo{author}{Min, B.} \&
  \bibinfo{author}{Goh, K.-I.}
\newblock \bibinfo{journal}{\bibinfo{title}{{Towards real-world complexity: an
  introduction to multiplex networks}}}.
\newblock {\emph{\JournalTitle{Eur. Phys. J. B}}}
  \textbf{\bibinfo{volume}{88}}, \bibinfo{pages}{48} (\bibinfo{year}{2015}).

\bibitem{Holme:2011fk}
\bibinfo{author}{Holme, P.} \& \bibinfo{author}{Saram{\"a}ki, J.}
\newblock \bibinfo{journal}{\bibinfo{title}{Temporal networks}}.
\newblock {\emph{\JournalTitle{Physics Reports}}}
  \textbf{\bibinfo{volume}{519}}, \bibinfo{pages}{97--125}
  (\bibinfo{year}{2012}).

\bibitem{Holme2015}
\bibinfo{author}{Holme, P.}
\newblock \bibinfo{journal}{\bibinfo{title}{{Modern temporal network theory: a
  colloquium}}}.
\newblock {\emph{\JournalTitle{Eur. Phys. J. B}}}
  \textbf{\bibinfo{volume}{88}}, \bibinfo{pages}{234} (\bibinfo{year}{2015}).

\bibitem{De-Domenico:2016aa}
\bibinfo{author}{De~Domenico, M.}, \bibinfo{author}{Granell, C.},
  \bibinfo{author}{Porter, M.~A.} \& \bibinfo{author}{Arenas, A.}
\newblock \bibinfo{journal}{\bibinfo{title}{The physics of spreading processes
  in multilayer networks}}.
\newblock {\emph{\JournalTitle{Nat Phys}}} \textbf{\bibinfo{volume}{12}},
  \bibinfo{pages}{901--906} (\bibinfo{year}{2016}).

\bibitem{citeulike:13204123}
\bibinfo{author}{De~Domenico, M.}, \bibinfo{author}{Sol\'{e}-Ribalta, A.},
  \bibinfo{author}{G\'{o}mez, S.} \& \bibinfo{author}{Arenas, A.}
\newblock \bibinfo{journal}{\bibinfo{title}{{Navigability of interconnected
  networks under random failures}}}.
\newblock {\emph{\JournalTitle{Proceedings of the National Academy of
  Sciences}}} \textbf{\bibinfo{volume}{111}}, \bibinfo{pages}{8351--8356}
  (\bibinfo{year}{2014}).

\bibitem{Buldyrev2010}
\bibinfo{author}{Buldyrev, S.~V.}, \bibinfo{author}{Parshani, R.},
  \bibinfo{author}{Paul, G.}, \bibinfo{author}{Stanley, H.~E.} \&
  \bibinfo{author}{Havlin, S.}
\newblock \bibinfo{journal}{\bibinfo{title}{{Catastrophic cascade of failures
  in interdependent networks}}}.
\newblock {\emph{\JournalTitle{Nature}}} \textbf{\bibinfo{volume}{464}},
  \bibinfo{pages}{1025--1028} (\bibinfo{year}{2010}).

\bibitem{6517108}
\bibinfo{author}{Yagan, O.}, \bibinfo{author}{Qian, D.},
  \bibinfo{author}{Zhang, J.} \& \bibinfo{author}{Cochran, D.}
\newblock \bibinfo{journal}{\bibinfo{title}{Conjoining speeds up information
  diffusion in overlaying social-physical networks}}.
\newblock {\emph{\JournalTitle{Selected Areas in Communications, IEEE Journal
  on}}} \textbf{\bibinfo{volume}{31}}, \bibinfo{pages}{1038--1048}
  (\bibinfo{year}{2013}).

\bibitem{Dickison2012}
\bibinfo{author}{Dickison, M.}, \bibinfo{author}{Havlin, S.} \&
  \bibinfo{author}{Stanley, H.~E.}
\newblock \bibinfo{journal}{\bibinfo{title}{Epidemics on interconnected
  networks}}.
\newblock {\emph{\JournalTitle{Physical Review E}}}
  \textbf{\bibinfo{volume}{85}}, \bibinfo{pages}{066109}
  (\bibinfo{year}{2012}).

\bibitem{barabasi2005origin}
\bibinfo{author}{Barab\'asi, A.}
\newblock \bibinfo{journal}{\bibinfo{title}{The origin of bursts and heavy
  tails in human dynamics}}.
\newblock {\emph{\JournalTitle{Nature}}} \textbf{\bibinfo{volume}{435}},
  \bibinfo{pages}{207--211} (\bibinfo{year}{2005}).

\bibitem{Stehle:2011}
\bibinfo{author}{Stehl\'e, J.} \emph{et~al.}
\newblock \bibinfo{journal}{\bibinfo{title}{High-resolution measurements of
  face-to-face contact patterns in a primary school}}.
\newblock {\emph{\JournalTitle{PLoS ONE}}} \textbf{\bibinfo{volume}{6}},
  \bibinfo{pages}{e23176} (\bibinfo{year}{2011}).

\bibitem{dynnetkaski2011}
\bibinfo{author}{Kivela, M.} \emph{et~al.}
\newblock \bibinfo{journal}{\bibinfo{title}{Multiscale analysis of spreading in
  a large communication network}}.
\newblock {\emph{\JournalTitle{J. Stat. Mech.}}} \bibinfo{pages}{P03005}
  (\bibinfo{year}{2012}).

\bibitem{PhysRevLett.98.158702}
\bibinfo{author}{Vazquez, A.}, \bibinfo{author}{R\'acz, B.},
  \bibinfo{author}{Luk\'acs, A.} \& \bibinfo{author}{Barab\'asi, A.-L.}
\newblock \bibinfo{journal}{\bibinfo{title}{Impact of non-poissonian activity
  patterns on spreading processes}}.
\newblock {\emph{\JournalTitle{Phys. Rev. Lett.}}}
  \textbf{\bibinfo{volume}{98}}, \bibinfo{pages}{158702}
  (\bibinfo{year}{2007}).

\bibitem{Parshani:2010}
\bibinfo{author}{Parshani, R.}, \bibinfo{author}{Dickison, M.},
  \bibinfo{author}{Cohen, R.}, \bibinfo{author}{Stanley, H.~E.} \&
  \bibinfo{author}{Havlin, S.}
\newblock \bibinfo{journal}{\bibinfo{title}{Dynamic networks and directed
  percolation}}.
\newblock {\emph{\JournalTitle{Europhysics Letters}}}
  \textbf{\bibinfo{volume}{90}}, \bibinfo{pages}{38004} (\bibinfo{year}{2010}).

\bibitem{PhysRevE.94.022316}
\bibinfo{author}{Moinet, A.}, \bibinfo{author}{Starnini, M.} \&
  \bibinfo{author}{Pastor-Satorras, R.}
\newblock \bibinfo{journal}{\bibinfo{title}{Aging and percolation dynamics in a
  non-poissonian temporal network model}}.
\newblock {\emph{\JournalTitle{Phys. Rev. E}}} \textbf{\bibinfo{volume}{94}},
  \bibinfo{pages}{022316} (\bibinfo{year}{2016}).

\bibitem{Jackson2010}
\bibinfo{author}{Jackson, M.}
\newblock \emph{\bibinfo{title}{Social and Economic Networks}}
  (\bibinfo{publisher}{Princeton University Press},
  \bibinfo{address}{Princeton}, \bibinfo{year}{2010}).

\bibitem{Verbrugge01061979}
\bibinfo{author}{Verbrugge, L.~M.}
\newblock \bibinfo{journal}{\bibinfo{title}{Multiplexity in adult
  friendships}}.
\newblock {\emph{\JournalTitle{Social Forces}}} \textbf{\bibinfo{volume}{57}},
  \bibinfo{pages}{1286--1309} (\bibinfo{year}{1979}).

\bibitem{Vijayaraghavan}
\bibinfo{author}{Vijayaraghavan, V.~S.}, \bibinfo{author}{No{\"e}l, P.-A.},
  \bibinfo{author}{Maoz, Z.} \& \bibinfo{author}{D'Souza, R.~M.}
\newblock \bibinfo{journal}{\bibinfo{title}{Quantifying dynamical spillover in
  co-evolving multiplex networks}}.
\newblock {\emph{\JournalTitle{Scientific Rep.}}}
  \textbf{\bibinfo{volume}{5}}, \bibinfo{pages}{15142 EP --}
  (\bibinfo{year}{2015}).

\bibitem{PhysRevLett.111.058702}
\bibinfo{author}{Kim, J.~Y.} \& \bibinfo{author}{Goh, K.-I.}
\newblock \bibinfo{journal}{\bibinfo{title}{Coevolution and correlated
  multiplexity in multiplex networks}}.
\newblock {\emph{\JournalTitle{Phys. Rev. Lett.}}}
  \textbf{\bibinfo{volume}{111}}, \bibinfo{pages}{058702}
  (\bibinfo{year}{2013}).

\bibitem{PhysRevLett.111.058701}
\bibinfo{author}{Nicosia, V.}, \bibinfo{author}{Bianconi, G.},
  \bibinfo{author}{Latora, V.} \& \bibinfo{author}{Barthelemy, M.}
\newblock \bibinfo{journal}{\bibinfo{title}{Growing multiplex networks}}.
\newblock {\emph{\JournalTitle{Phys. Rev. Lett.}}}
  \textbf{\bibinfo{volume}{111}}, \bibinfo{pages}{058701}
  (\bibinfo{year}{2013}).

\bibitem{Karsai:2012aa}
\bibinfo{author}{Karsai, M.}, \bibinfo{author}{Kaski, K.},
  \bibinfo{author}{Barab{\'a}si, A.-L.} \& \bibinfo{author}{Kert{\'e}sz, J.}
\newblock \bibinfo{journal}{\bibinfo{title}{Universal features of correlated
  bursty behaviour}}.
\newblock {\emph{\JournalTitle{Sci. Rep.}}} \textbf{\bibinfo{volume}{2}}
  (\bibinfo{year}{2012}).

\bibitem{PhysRevE.92.022814}
\bibinfo{author}{Jo, H.-H.}, \bibinfo{author}{Perotti, J.~I.},
  \bibinfo{author}{Kaski, K.} \& \bibinfo{author}{Kert\'esz, J.}
\newblock \bibinfo{journal}{\bibinfo{title}{Correlated bursts and the role of
  memory range}}.
\newblock {\emph{\JournalTitle{Phys. Rev. E}}} \textbf{\bibinfo{volume}{92}},
  \bibinfo{pages}{022814} (\bibinfo{year}{2015}).

\bibitem{Granell2013}
\bibinfo{author}{Granell, C.}, \bibinfo{author}{G\'omez, S.} \&
  \bibinfo{author}{Arenas, A.}
\newblock \bibinfo{journal}{\bibinfo{title}{Dynamical interplay between
  awareness and epidemic spreading in multiplex networks}}.
\newblock {\emph{\JournalTitle{Phys. Rev. Lett.}}}
  \textbf{\bibinfo{volume}{111}}, \bibinfo{pages}{128701}
  (\bibinfo{year}{2013}).

\bibitem{eagle2006reality}
\bibinfo{author}{Eagle, N.} \& \bibinfo{author}{Pentland, A.}
\newblock \bibinfo{journal}{\bibinfo{title}{Reality mining: sensing complex
  social systems}}.
\newblock {\emph{\JournalTitle{Personal and Ubiquitous Computing}}}
  \textbf{\bibinfo{volume}{10}}, \bibinfo{pages}{255--268}
  (\bibinfo{year}{2006}).

\bibitem{PhysRevE.91.052813}
\bibinfo{author}{Xuan, Q.}, \bibinfo{author}{Fang, H.}, \bibinfo{author}{Fu,
  C.} \& \bibinfo{author}{Filkov, V.}
\newblock \bibinfo{journal}{\bibinfo{title}{Temporal motifs reveal
  collaboration patterns in online task-oriented networks}}.
\newblock {\emph{\JournalTitle{Phys. Rev. E}}} \textbf{\bibinfo{volume}{91}},
  \bibinfo{pages}{052813} (\bibinfo{year}{2015}).

\bibitem{Apache}
\bibinfo{title}{http://www.apache.org/}.
\newblock \urlprefix\url{http://www.apache.org/}.

\bibitem{newmancitations01}
\bibinfo{author}{Newman, M. E.~J.}
\newblock \bibinfo{journal}{\bibinfo{title}{The structure of scientific
  collaboration networks}}.
\newblock {\emph{\JournalTitle{Proc. Natl. Acad. Sci. USA}}}
  \textbf{\bibinfo{volume}{98}}, \bibinfo{pages}{404--409}
  (\bibinfo{year}{2001}).

\bibitem{apsdata}
\bibinfo{title}{{American Physical Society. Data sets for research}}.
\newblock \urlprefix\url{https://publish.aps.org/datasets}.

\bibitem{Song1018}
\bibinfo{author}{Song, C.}, \bibinfo{author}{Qu, Z.}, \bibinfo{author}{Blumm,
  N.} \& \bibinfo{author}{Barab{\'a}si, A.-L.}
\newblock \bibinfo{journal}{\bibinfo{title}{Limits of predictability in human
  mobility}}.
\newblock {\emph{\JournalTitle{Science}}} \textbf{\bibinfo{volume}{327}},
  \bibinfo{pages}{1018--1021} (\bibinfo{year}{2010}).

\bibitem{PhysRevX.1.011008}
\bibinfo{author}{Takaguchi, T.}, \bibinfo{author}{Nakamura, M.},
  \bibinfo{author}{Sato, N.}, \bibinfo{author}{Yano, K.} \&
  \bibinfo{author}{Masuda, N.}
\newblock \bibinfo{journal}{\bibinfo{title}{Predictability of conversation
  partners}}.
\newblock {\emph{\JournalTitle{Phys. Rev. X}}} \textbf{\bibinfo{volume}{1}},
  \bibinfo{pages}{011008} (\bibinfo{year}{2011}).

\bibitem{mobilitypetri2012}
\bibinfo{author}{Szell, M.}, \bibinfo{author}{Sinatra, R.},
  \bibinfo{author}{Petri, G.}, \bibinfo{author}{Thurner, S.} \&
  \bibinfo{author}{Latora, V.}
\newblock \bibinfo{journal}{\bibinfo{title}{Understanding mobility in a social
  petri dish}}.
\newblock {\emph{\JournalTitle{Sci. Rep.}}} \textbf{\bibinfo{volume}{2}},
  \bibinfo{pages}{457} (\bibinfo{year}{2012}).

\bibitem{Bullmore2012}
\bibinfo{author}{Bullmore, E.} \& \bibinfo{author}{Sporns, O.}
\newblock \bibinfo{journal}{\bibinfo{title}{The economy of brain network
  organization}}.
\newblock {\emph{\JournalTitle{Nature Reviews Neuroscience}}}
  \textbf{\bibinfo{volume}{13}}, \bibinfo{pages}{336--349}
  (\bibinfo{year}{2012}).

\bibitem{Gallotti:2014aa}
\bibinfo{author}{Gallotti, R.} \& \bibinfo{author}{Barthelemy, M.}
\newblock \bibinfo{journal}{\bibinfo{title}{Anatomy and efficiency of urban
  multimodal mobility}}.
\newblock {\emph{\JournalTitle{Scientific Rep.}}}
  \textbf{\bibinfo{volume}{4}}, \bibinfo{pages}{6911} (\bibinfo{year}{2014}).

\bibitem{Gallottie1500445}
\bibinfo{author}{Gallotti, R.}, \bibinfo{author}{Porter, M.~A.} \&
  \bibinfo{author}{Barthelemy, M.}
\newblock \bibinfo{journal}{\bibinfo{title}{Lost in transportation: Information
  measures and cognitive limits in multilayer navigation}}.
\newblock {\emph{\JournalTitle{Science Advances}}}
  \textbf{\bibinfo{volume}{2}}, \bibinfo{pages}{1500445}
  (\bibinfo{year}{2016}).

\bibitem{Lee:2010fk}
\bibinfo{author}{Lee, S.}, \bibinfo{author}{Rocha, L. E.~C.},
  \bibinfo{author}{Liljeros, F.} \& \bibinfo{author}{Holme, P.}
\newblock \bibinfo{journal}{\bibinfo{title}{Exploiting temporal network
  structures of human interaction to effectively immunize populations}}.
\newblock {\emph{\JournalTitle{PLoS ONE}}} \textbf{\bibinfo{volume}{7}},
  \bibinfo{pages}{e36439} (\bibinfo{year}{2012}).

\bibitem{2013arXiv1305.2357S}
\bibinfo{author}{{Starnini}, M.}, \bibinfo{author}{{Machens}, A.},
  \bibinfo{author}{{Cattuto}, C.}, \bibinfo{author}{{Barrat}, A.} \&
  \bibinfo{author}{{Pastor Satorras}, R.}
\newblock \bibinfo{journal}{\bibinfo{title}{{Immunization strategies for
  epidemic processes in time-varying contact networks}}}.
\newblock {\emph{\JournalTitle{Journal of Theoretical Biology}}}
  \textbf{\bibinfo{volume}{337}}, \bibinfo{pages}{89--100}
  (\bibinfo{year}{2013}).

\bibitem{cox1980point}
\bibinfo{author}{Cox, D.} \& \bibinfo{author}{Isham, V.}
\newblock \emph{\bibinfo{title}{Point Processes}}.
\newblock Chapman \& Hall/CRC Monographs on Statistics \& Applied Probability
  (\bibinfo{publisher}{Taylor \& Francis}, \bibinfo{address}{Cambridge, U.K.},
  \bibinfo{year}{1980}).

\bibitem{renewal}
\bibinfo{author}{Cox, D.~R.}
\newblock \emph{\bibinfo{title}{Renewal Theory}} (\bibinfo{publisher}{Methuen},
  \bibinfo{address}{London}, \bibinfo{year}{1967}).

\end{thebibliography}
\end{document}